\documentstyle[twocolumn,prb,aps,psfig]{revtex}
\newcommand{\firma}[1]{\texttt{#1}}
\def\be{\begin{equation}}
\def\ee{\end{equation}}
\def\bc{\begin{center}}
\def\ec{\end{center}}
\begin{document}
\draft
\title{Ising films with surface defects}
\author{M.-C.\ Chung \dag\ , M. Kaulke \dag\ \thanks{\firma{Now at: I-D Media AG,
Weidenallee 37a , D-20357 Hamburg, Germany}} , I. Peschel \dag\ , M. Pleimling \ddag\ and W. Selke \$ }
\address{
\dag Fachbereich Physik, Freie Universi\"at Berlin, D--14195 Berlin, Germany\\
\ddag Institut f\"ur Theoretische Physik I, Universit\"at Erlangen-N\"urnberg,
D--91058 Erlangen, Germany\\
\$ Institut f\"ur Theoretische Physik, Technische Hochschule, D--52056 Aachen, Germany }
\maketitle
 
\begin{abstract}
The influence of surface defects on the critical properties of
magnetic films is 
studied for Ising models with nearest-neighbour ferromagnetic
couplings. The defects include one or two adjacent lines of
additional atoms and
a step on the surface. For the calculations, both density-matrix 
renormalization group and Monte Carlo techniques are used. 
By changing the local couplings at the defects and the film
thickness, non-universal features as well as interesting
crossover phenomena in the magnetic exponents are observed.
\end{abstract}

 
\section{Introduction}
Critical phenomena of magnetic films are of current
interest, both experimentally and
theoretically \cite{Gra,Sch98,Qui94,Jan93,Mar00}. In the limiting
cases of one layer
and of infinitely many layers, one deals with two-dimensional magnets 
\cite{Igl93} and with standard bulk and surface magnetism \cite{Bin83,Die86,Dos92,Kan91}, respectively. For systems consisting of a finite number of
layers, interesting crossover phenomena between these
limiting cases are expected.  

In this article, we shall consider critical properties of 
ferromagnetic films of Ising magnets
with various imperfections at the surface, motivated partly
by possible experimental realizations 
of magnetic thin films with stripes of magnetic
adatoms and stepped surfaces \cite{Gra,Kirsch,Kirsch2}, partly
by genuine theoretical interest. Imperfections may
be due to regular or irregular changes
in the surface couplings or due to additional structures on
the surface. A simple example of the first case is a ladder of modified 
couplings in an otherwise uniform two-dimensional system as
introduced
by Bariev \cite{Bar79}, see Fig. 1(a). We will study this 
briefly, since
it can serve as a testing ground. Our main interest, however, is in
additional structures, as depicted in Figs. 1(b)--(d).
Thus we will investigate surfaces with magnetic adatoms in the form of
\begin{itemize}
\item one additional straight line, Fig. 1(b),
\item two neighbouring lines, Fig. 1(c),
\item a straight step of unit height, Fig. 1(d),
\end{itemize}
for various local couplings at the defects and for films of
varying thickness.

\begin{figure}
\centerline{\psfig{figure=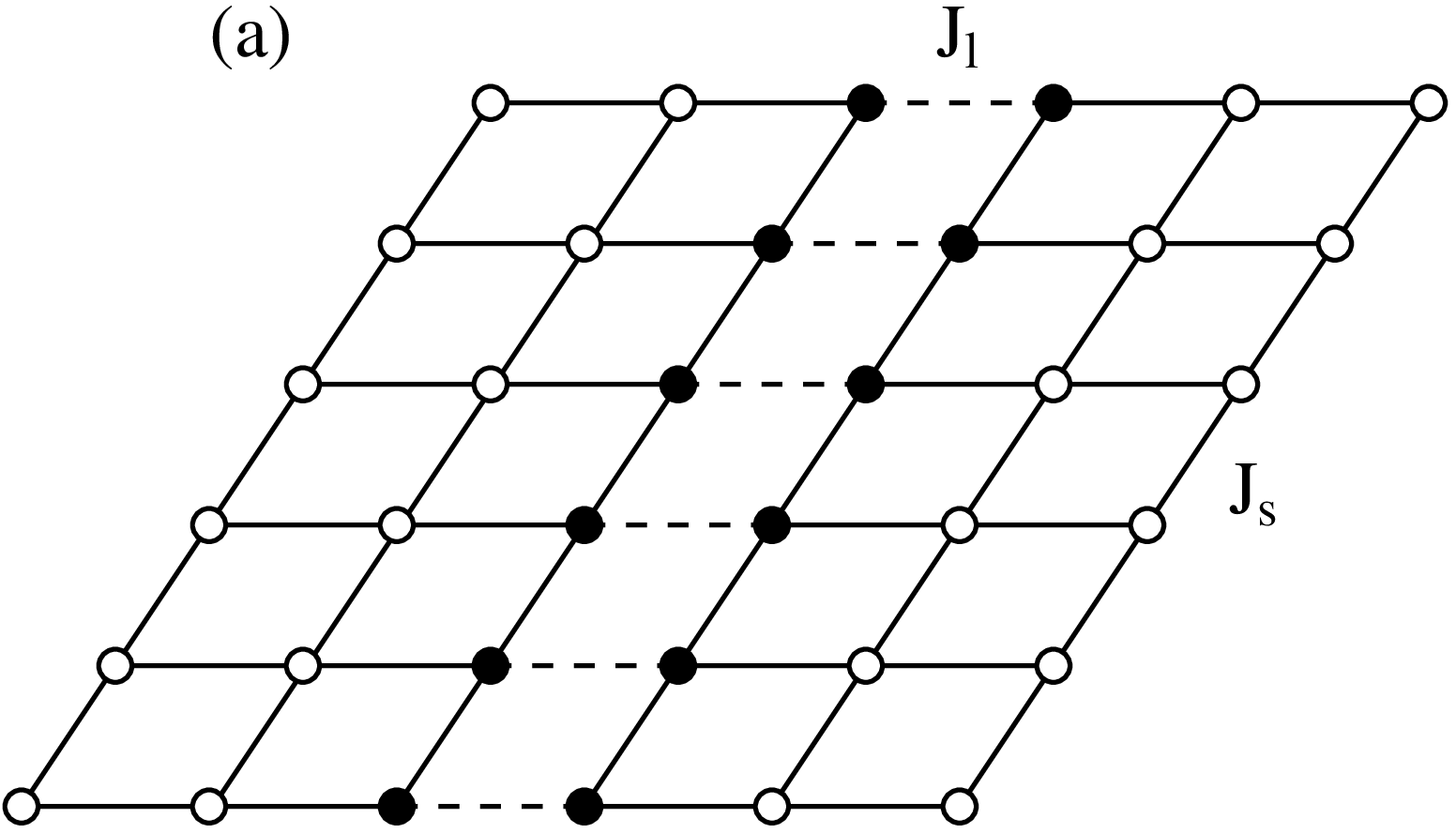,width=2.75in,angle=270}}
\vspace*{0.5cm}
\centerline{\psfig{figure=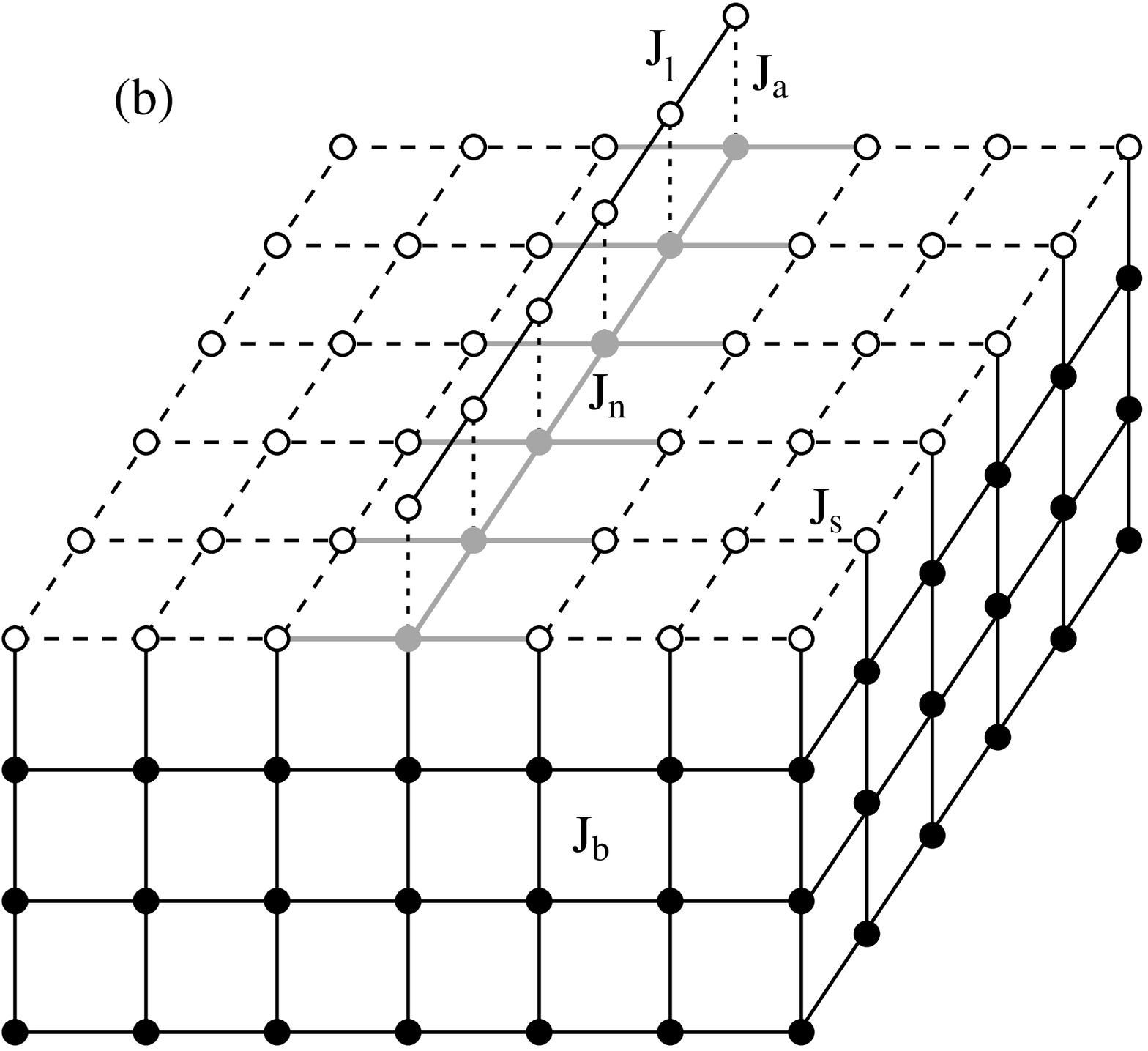,width=2.75in,angle=270}}
\vspace*{0.5cm}
\centerline{\psfig{figure=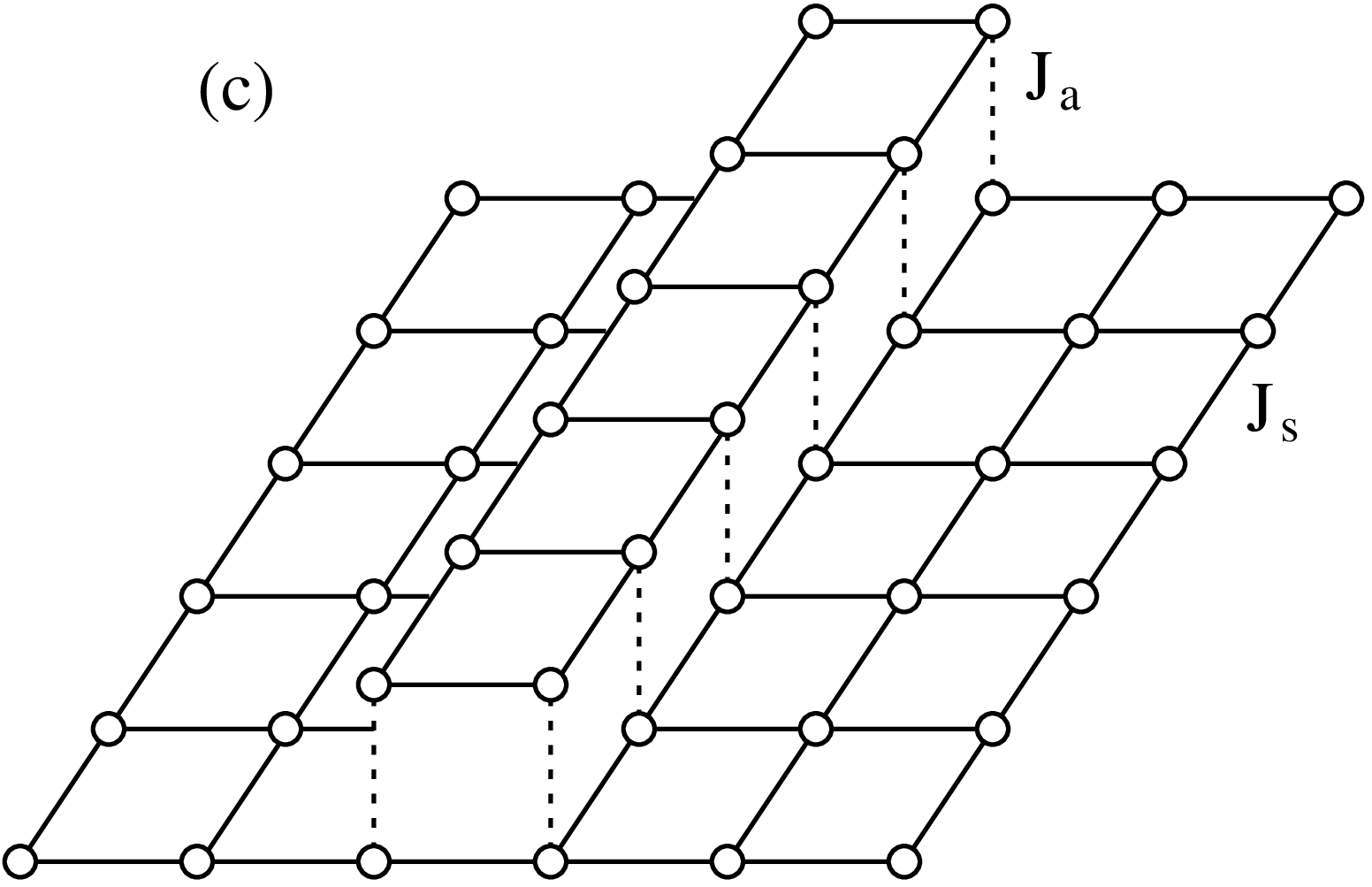,width=2.75in,angle=270}}
\centerline{\psfig{figure=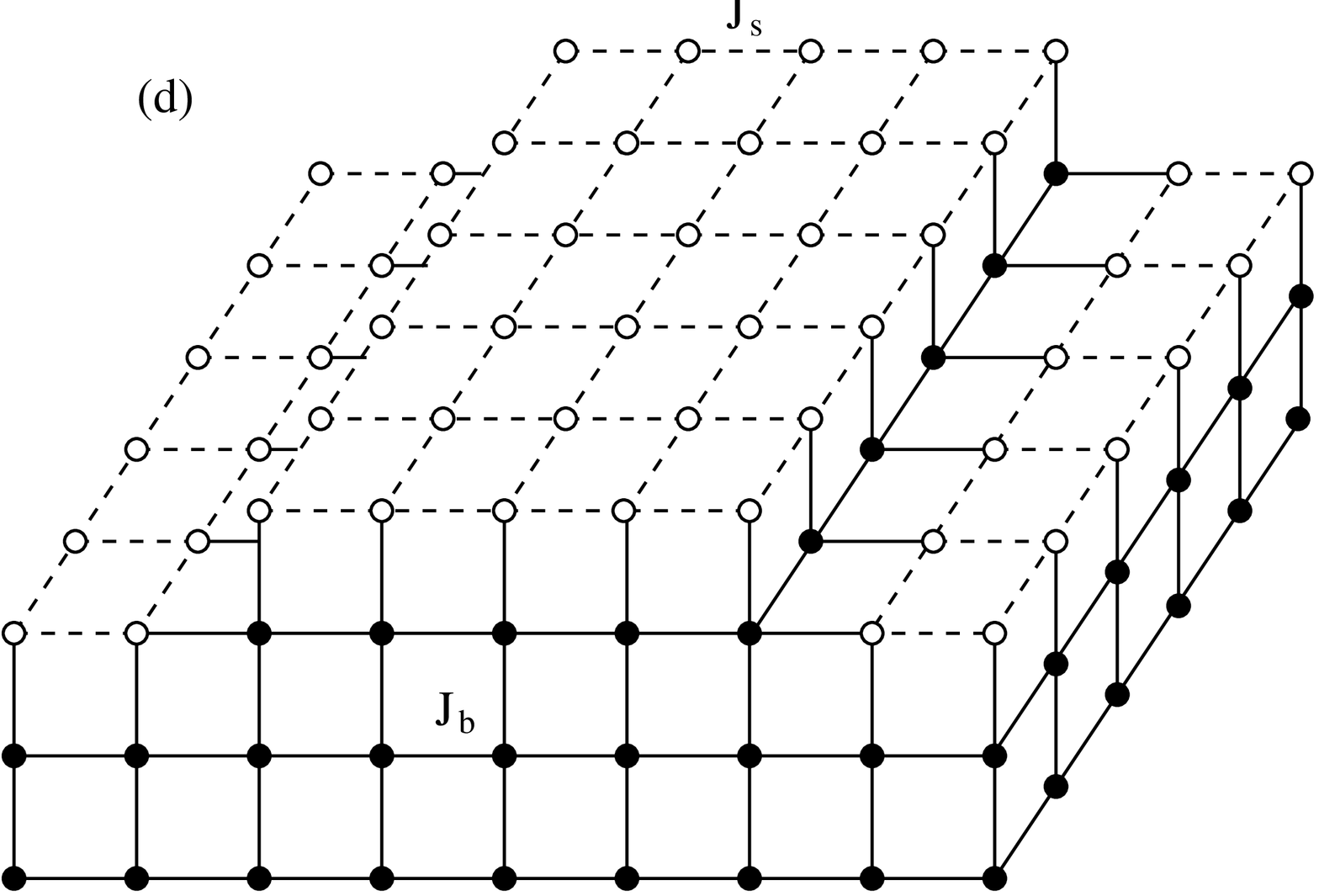,width=2.75in,angle=270}}
\vspace*{0.5cm}
\caption{Geometry and interactions of Ising models 
with various surface imperfections: (a) ladder defect, (b)
one additional line of spins, (c) pair of adjacent lines of magnetic
adatoms, and (d) straight step of monoatomic height. Usually, surface
spins are denoted by open circles, and bulk spins by full circles. In
case 1(b), the shaded spins allow for both interpretations.}
\label{fig1} \end{figure}

Previous related work on Ising models includes the study of the
step magnetization at the ordinary transition of rather thick
films \cite{Ple98} and the study of magnetism in thin films with 
rough surfaces \cite{Rei98}.

In studying the influence of these
imperfections especially on the critical behaviour, we use the density-
matrix renormalization group technique (DMRG) \cite{Whi92,Pes99}, being
most suitable in the case of merely one layer, and the Monte
Carlo (MC) method \cite{Bin92}, which allows to treat films of considerable
thickness as well.
  
The article is organized as follows. In the next section, we present
our findings on single layers with defects, applying DMRG. The MC
results on single layers and on films with an additional
line of magnetic adatoms and with a straight step on top of
the surface are discussed in Section 3. A short summary concludes
the article.

\section{One layer: DMRG}

The planar Ising model with line-like defects is a peculiar system,
because it shows non-universal magnetic exponents. This is
connected with the values $\nu =1$ and $x_s = 1/2$ of the exponents
for the correlation length and the surface magnetization of the pure
system, respectively. A one-dimensional, energy-like perturbation then 
is marginal and can change the critical behaviour continuously. For this 
reason, the system has been the topic of various studies \cite{Igl93}, 
with the focus most recently on a conformal treatment \cite{Osh96} 
and on random systems \cite{Sza99}.
While the simple chain and ladder defects considered by Bariev are
solvable free-fermion problems, the other cases we study are not integrable
and one has to use numerical methods. In the following we dicuss the
quantity of direct physical interest, the local magnetization at or near
the defect lines.

To obtain it, we used the transfer matrix running along the direction
of the defect, see Figs. 1(a)--(c), and determined its maximal eigenvector
via the DMRG method \cite{Nis99}. In this way one is treating an infinitely
long strip of width $M$ with the defect located in the middle.
Only the infinite-system algorithm was used, in which one enlarges the
system step by step and always chooses an optimal reduced basis 
via the density matrix. This is very
convenient, since one can insert different defects after the 
system has reached the desired size. No further sweeps to optimize the
state were made, since tests on the ladder defect gave very good results
without them. Most calculations were done with $64$ kept states and a
truncation error around $10^{-15}$.
The local magnetization $m(i)$ was determined from the spin correlation
function $C(i) = \langle\sigma_1 \sigma_i\rangle$
 for free boundary conditions, or 
directly as $\langle\sigma_i\rangle$
  for fixed boundary spins. The width was always
much larger than the correlation length and 
varied between $M=100$ and $M=5000$ 
for the temperature range studied ($0.001 < t < 0.1$, where $t=1-T/T_c$ is the 
reduced temperature). The (absolute) error in $m$, 
determined by comparing with 
anaytical results was at most $10^{-4}$  for a system at $t =0.001$,
 cut in the 
middle by a ladder defect. 
For less severe modifications and larger values of $t$
it was even smaller. 

\begin{figure}
\centerline{\psfig{figure=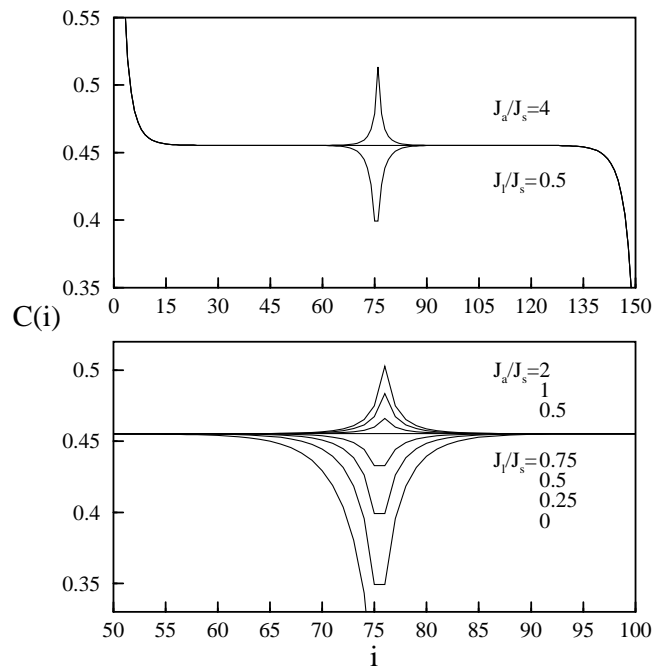,width=3.35in}}
\vspace*{0.2cm}
\caption{Spin correlation function $C(i)$ for a strip of width $M=150$ 
with ladder defects
(below the plateau) or one additional line of spins (above the plateau), as
obtained from DMRG calculations at
the reduced temperature $t=0.072$. The defect
strengths $J_l/J_s$ and $J_a/J_s$ are 
indicated. Upper part: total view, lower part: central region.}
\label{fig2} \end{figure}

In Fig. 2 we show the correlation function $C(i)$ across the
strip for ladder defects (Fig.\ 1(a)) and for 
an additional line (Fig.\ 1(b); in the
DMRG study we considered the case $J_n= J_s$). The upper part gives
an overall picture, while the lower one shows the defect region in
more detail. For ladder defects the strength $J_l$ of the
defect bonds was varied, whereas for an additional line it was 
the coupling $J_a$ between the line spins and the substrate.
Since $C(i)$ factors for large distances, these curves also give
the profile of the magnetization in the bulk. 
One can see how $m$ increases or decreases 
near the defect, depending on the sign of the perturbation (similar curves 
were obtained in \cite{Sza99} for a random system).
If one cuts the ladder bonds by choosing $J_l = 0$,
one obtains the boundary magnetization of the homogeneous model in
the middle of the strip. On the upper side, the possible increase 
of $m$ depends on the details
of the defect. It is limited if one varies $J_a$, because a line with
infinite $J_a$ is equivalent to a chain defect in the plane with
merely doubled bond strength.

The temperature dependence of $m$ is shown in Fig. 3 for the spins in 
the plane situated below one or two additional lines. One can see
how it is increased over the Onsager value by increasing the coupling $J_a$.
As expected, the effect is even stronger for two additional lines.
In this case, $m$ has already twice the undisturbed value for the smallest
shown $t$. Quantitatively, this enhancement is described by a decrease of the 
exponent $\beta_l$, the local critical exponent which describes the vanishing
of the magnetization near the additional line of magnetic adatoms. 

\begin{figure}
\centerline{\psfig{figure=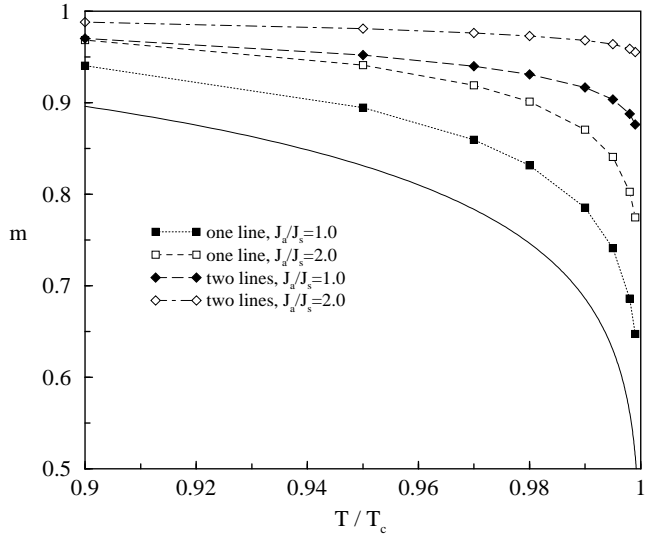,width=3.35in,angle=270}}
\caption{Local magnetization $m$ of the spins below one or two additional
lines as a function of temperature, for three values of the
coupling ratio $J_a/J_s$. The lowest curve is the Onsager result for
the perfect Ising model.}
\label{fig3} \end{figure}
  
To investigate this, we have analyzed the temperature behaviour of $m$
in terms of an effective (critical) exponent $\beta_{eff}$, defined
by\cite{Sch98,Ple98,Ple99}
\begin{equation}
          \beta_{eff}(t)\; =\;\ln(m(t_i)/m(t_{i+1}))/\ln(t_i/t_{i+1})  
\end{equation} 
with $t=(t_i+t_{i+1})/2$  (alternatively, one could choose $t$ to be the geometric mean
$t=\sqrt{t_it_{i+1}}$). As one approaches the critical point, $t \rightarrow 0$,
this quantity converges to the true local exponent $\beta_l$. It is also
a very sensitive indicator for the numerical accuracy of a calculation.  

\begin{figure}        
\centerline{\psfig{figure=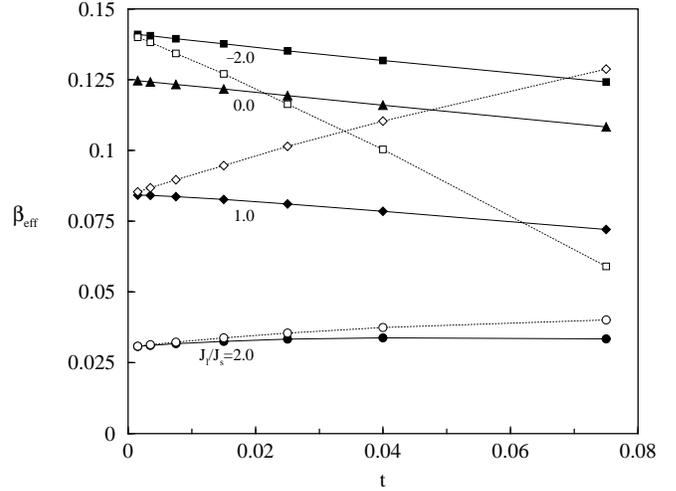,width=3.35in,angle=270}}
\caption{Effective exponent $\beta_{eff}$ as function of the
reduced temperature $t$
for one additional line and three different coupling ratios $J_l/J_s$.
Full: Spins located below the line, dotted: spins in the line.}
\label{fig4} \end{figure}
         
Some typical results are given in Fig. 4 for one additional line and
four values of the ratio $J_l/J_s$ of the couplings
in the line. For $J_l$ = 0, 
one is treating the plane with independent attached spins and the 
Onsager result $\beta = 0.125$ is recovered with high accuracy. 
In the other cases, the exponents both
for the spin in the line and the one below it are shown 
and one sees that the two curves have different slopes, 
but a common limit for $t \rightarrow 0$
 which can be determined very precisely. 
The values for $\beta_l$ found in this way are accurate to at 
least three digits. For the case $J_a/J_s \gg 1$ which, as mentioned, 
is equivalent to a line defect in the plane, this was
checked explicitely by comparing with the analytical result. In the
figure, also a negative $J_l$ is shown, which leads to a reduction of $m$
and an increase of $\beta_l$ over the Onsager value . In this case, a
limiting value $0.142$  is approached rapidly for $J_l/J_s < -1$.
This is the same effect as for a chain defect in the plane with strong 
antiferromagnetic couplings \cite{Igl93}. 
In that case, the exponent is increased up to the value 0.5 of the free surface.
The sign of $J_a$, on the other hand, has no influence on the exponent. 

The results for $\beta_l$ are collected in Table 1 and in Fig. 5, 
where the exponent is plotted as a function of the 
varied couplings (keeping
the other couplings fixed and equal
to $J_s$). For comparison also the analytical results
\cite{Bar79,Igl93}, for
simple chain and ladder defects are shown
in Fig. 5. One notes that, for a single 
line and small modifications, it does not matter much whether 
one changes $J_a$ or $J_l$. A large $J_l/J_s$, however, has a much more 
pronounced effect than $J_a/J_s$, since it corresponds to additional spins 
which are almost rigidly locked together. For the double line, 
the exponent drops much faster,  
reaching $10^{-2}$ already around $J_a/J_s \sim 1$. 
For more additional lines, i.e. for a terrace on the surface as in Fig. 1(d), 
this effect would be even stronger. In this case, 
the magnetization would practically
jump as in a first-order transition.

\begin{figure}
\centerline{\psfig{figure=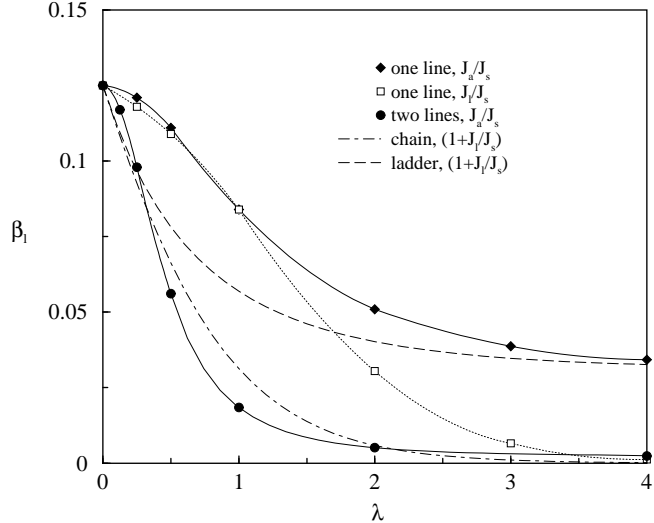,width=3.35in,angle=270}}
\caption{Local exponent $\beta_l$ as a function of the ratio of the coupling 
strength $\lambda$, as defined in the figure for five different
situations. For chain and ladder defects
the analytical results are shown, otherwise the DMRG results are depicted.}
\label{fig5} \end{figure}

\begin{table}          
\caption{Numerical values for the local exponent $\beta_l$ of an
Ising plane with one and two additional lines of spins.}              
\begin{tabular}{|c|c|c|c|} \hline
   \makebox[20mm]{lines} &\makebox[20mm]{1} &\makebox[20mm]{1} &
 \makebox[20mm]{2} \\
  \hline 
   $\lambda$ & $J_a/J_s$ & $J_l/J_s$ & $J_a/J_s$ \\ 
  \hline
  $ 0.0$ & $0.125$ & $0.125$ & $0.125$ \\
  \hline
 $0.25$  & $0.121$ & $0.118$ & $0.098$ \\
 \hline  
 $0.5$  & $0.111$ & $0.109$ & $0.056$ \\
 \hline  
 $1.0$  & $0.084$ & $0.084$ & $0.018$ \\
 \hline  
 $2.0$  & $0.051$ & $0.031$ & $0.005$ \\
 \hline  
 $4.0$  & $0.034$ & $0.002$ & $0.001$ \\
\end{tabular}
\label{table1} 
\end{table} 

\section{Films: Monte Carlo simulations}

\subsection{One additional line of spins}

Extending the DMRG calculations on an Ising layer with one 
additional line of spins, we did Monte
Carlo simulations on the corresponding Ising films, consisting
of $L$ layers with 
one line of magnetic adatoms on top of the surface, see
Fig. 1(b). We 
set $J_l= J_s$, with $J_a= J_n= J_s$ (variant A, treating the spins 
directly below the additional line as surface spins, as it was done
in the DMRG study) or
$J_a=J_n= J_b$ (variant B, treating the
spins directly below the additional line as bulk
spins).-- In the 
layers, periodic boundary conditions are
used.

Let $S_{i,j,k} = \pm 1$ be the spin on site $(i,j)$ in the
$k$th layer. Taking layers of $M \times N$ spins, the
spins in the additional line on top of the surface
through the center are located
at $(i=(M+1)/2,j,k=0)$, $M$ odd, with $j$ running from
1 to $N$. We computed, among others, the line magnetization $m(i,k;L)$ 
defined by  

\be
m(i,k;L)= \frac{1}{N} \left< \left| \sum\limits_j  S_{i,j,k} \right| \right>
\ee
%
\noindent
The magnetization of the line on top of the surface, $m_l$, is given
by $m_l(L)$= $m((M+1)/2,k=0;L)$.

In the simulations, the film thickness $L$ ranged from 1 to 40, with 
layer sizes being sufficiently large to circumvent finite--size effects (up
to $161 \times 320$). To speed up computations, the
single--cluster--flip algorithm was implemented. We studied
the cases (i) $J_s = J_b$ as well as (ii) $J_s = 2 J_b$ 
(variants A and B), which lead to the
two characteristic scenarios of surface critical
phenomena for $L, M, N \rightarrow \infty$ (semi--infinite case). 
In the first case, bulk and surface spins
order simultaneously at temperature $T_c$ (ordinary
transition), while in the second one the surface spins order at 
a higher temperature, $T_s$ (surface transition) \cite{Bin83,Die86}.

(i) At the {\it ordinary} transition of the semi--infinite Ising
model, $L \rightarrow \infty$, the
magnetization deep in the bulk vanishes like $m \propto t^{\beta}$, with
$t= |T- T_c|/T_c$, where $\beta= 0.31...$ \cite{Fer91,Tal96}. 
At the perfect, flat
surface, one finds $m \propto t^{\beta_1}$, with
$\beta_1 \approx 0.80$ \cite{Lan90,Ple98}. The vanishing of the 
magnetization in the additional line of spins on top of
the surface is expected to be governed by $\beta_1$ as well, i.e.
$\beta_l(L \rightarrow \infty)= \beta_1$ \cite{Ple98,Die90}.  On
the other hand, for a single perfect layer, $L= 1$,
it is well known that $m \propto t^{\beta_{2d}}$, $\beta_{2d}= 1/8$. Adding
a row of spins, we obtain, from
the DMRG calculations, $m_l \propto t^{\beta_l(L=1)}$ with    
$\beta_l(L=1) \approx 0.084$, see Table 1.

To monitor the influence of the layer thickness $L$ at the
ordinary transition, we
computed magnetization profiles $m(i,k;L)$, the critical
temperature $T_c(L)$, and the critical exponent $\beta_l(L)$. The dependence
of the transition temperature on the thickness $L$ has
been studied before for flat films \cite{Bin83,Rei98}, and it
is, certainly, not affected by the presence of the additional line.
In Fig.\ 6, the magnetization in the defect line, $m_l(L)$, is depicted
as a function of temperature for $L$ ranging from
1 to 5, illustrating the increase of the transition
temperature with $L$. In the
ordered phase, $T < T_c$, the line magnetization $m(i,k;L)$ is, in
each layer, maximal for the center line, $m((M+1)/2,k;L)$, see
also Fig.\ 2. The maximum is
most pronounced at $k=1$ (we shall denote the magnetization in
that line beneath the additional row of spins by $m_{lb}= 
m((M+1)/2,1;L)$), due to the increased coordination number, compared
to the other surface lines. The magnetization in the additional
line, $m_l$, is suppressed compared to $m_{lb}$, because of
missing neighbouring spins.

\begin{figure}
\centerline{\psfig{figure=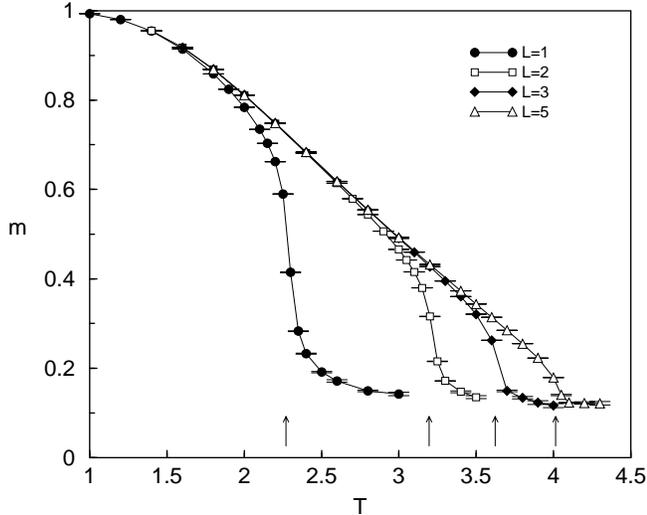,width=3.35in,angle=270}}
\caption{
Simulated magnetization in the additional line $m_l(L)$ for Ising
films with $L$ layers and one additional
line of spins on the surface, choosing $J_s= J_b$, for
$L$ ranging from 1 to 5. Each layer consists of $81 \times 80$
spins. The
critical temperatures $T_c(L)$ are marked by arrows.}
\label{fig6} \end{figure}

Various crossover effects show up in the effective exponent
$\beta_{eff} (i,k;L)$, defined
by $m(i,k;L) \propto t^{\beta_{eff}(i,k;L)}$, corresponding to
the slope in a standard log--log--plot of the temperature
dependence of the magnetization \cite{Sch98,Ple98,Ple99}, see
eq. (1). On approach
to $T_c$, one
expects to observe the limiting cases  
$\beta_{eff} (i,k;L) \rightarrow \beta$ for $k$ and $L$ large,
 $\rightarrow \beta_1$ for $k$ small and $L$
large, $\rightarrow \beta_l(L=1)$ for $L= 1$ and $i= (M+1)/2$, $k= 0$ or
1, and $\rightarrow \beta_{2d}$ for $L=1$ and sufficiently
far away from the additional line in the centre.

The crossover behaviour is illustrated in Fig. 7, showing
$\beta_{eff} ((M+1)/2,k;L)$ for $m_l(L)$, $k= 0$, and
$m_{lb}(L)$, $k= 1$, with
the film thickness ranging from $L= 1$ to
10. $\beta_{eff} ((M+1)/2,0;L)$ decreases monotonically, except for $L=1$, 
over a wide range of temperatures on lowering $t$, but with the effective
exponent, at fixed $t$, increasing clearly with the film
thickness, as depicted in Fig. 7(a). The data seem to indicate
that the asymptotic critical
exponent $\beta_l(L)$, as $t \rightarrow 0$, of the magnetization in the
additional line of
magnetic adatoms increases, however, only weakly with $L$, being quite
small, around 0.1, for $L$ going up to 10 (the increase itself
may be argued to reflect the diminishing role of the
defect line on the two--dimensional
critical fluctuations in thicker films; of course, $\beta(L)$
is bounded by 1/8 for finite $L$). For 
the magnetization beneath
the additional line, $m_{lb}$, corrections to the asymptotics are
rather large as well, see Fig.\ 7(b). 
Here the effective exponent
$\beta_{eff}((M+1)/2,1;L)$ changes with temperature in a non--monotonic
fashion, except for $L= 1$. In agreement with the observations for
$m_l$, the true critical exponent $\beta_l(L)$ 
is rather small, around 0.1, increasing only weakly with $L$. The location 
of the maximum in $\beta_{eff}((M+1)/2,1;L)$ indicates the temperature, at
which one crosses over from the regime dominated
by two--dimensional critical fluctuations, close
to the phase transition, to the
regime, further away from the critical point, where
the fluctuations are (nearly) isotropic and
three--dimensional. Thence, at the maximum the corresponding
correlation length is argued to be about the thickness of the
film $L$. In the thermodynamic limit, $L \rightarrow \infty$, the
maximum is believed to shift towards $t= 0$, with its height being
$\beta_l = \beta_1 \approx 0.80$.

Note that the strong corrections to scaling, as seen
by the deviations of the effective exponents from their
asymptotic values, may cause severe difficulties in extracting
the true critical exponents in simulations as well as in
experiments.

Similar crossover phenomena, now between $\beta_{2d}$, $\beta_1$ and
$\beta$, are expected to occur for the magnetization far away
from the defect line, when varying the
film thickness. This aspect, however, is of minor importance in
the context of this study.

\vspace*{1cm}
\begin{figure}
\centerline{\psfig{figure=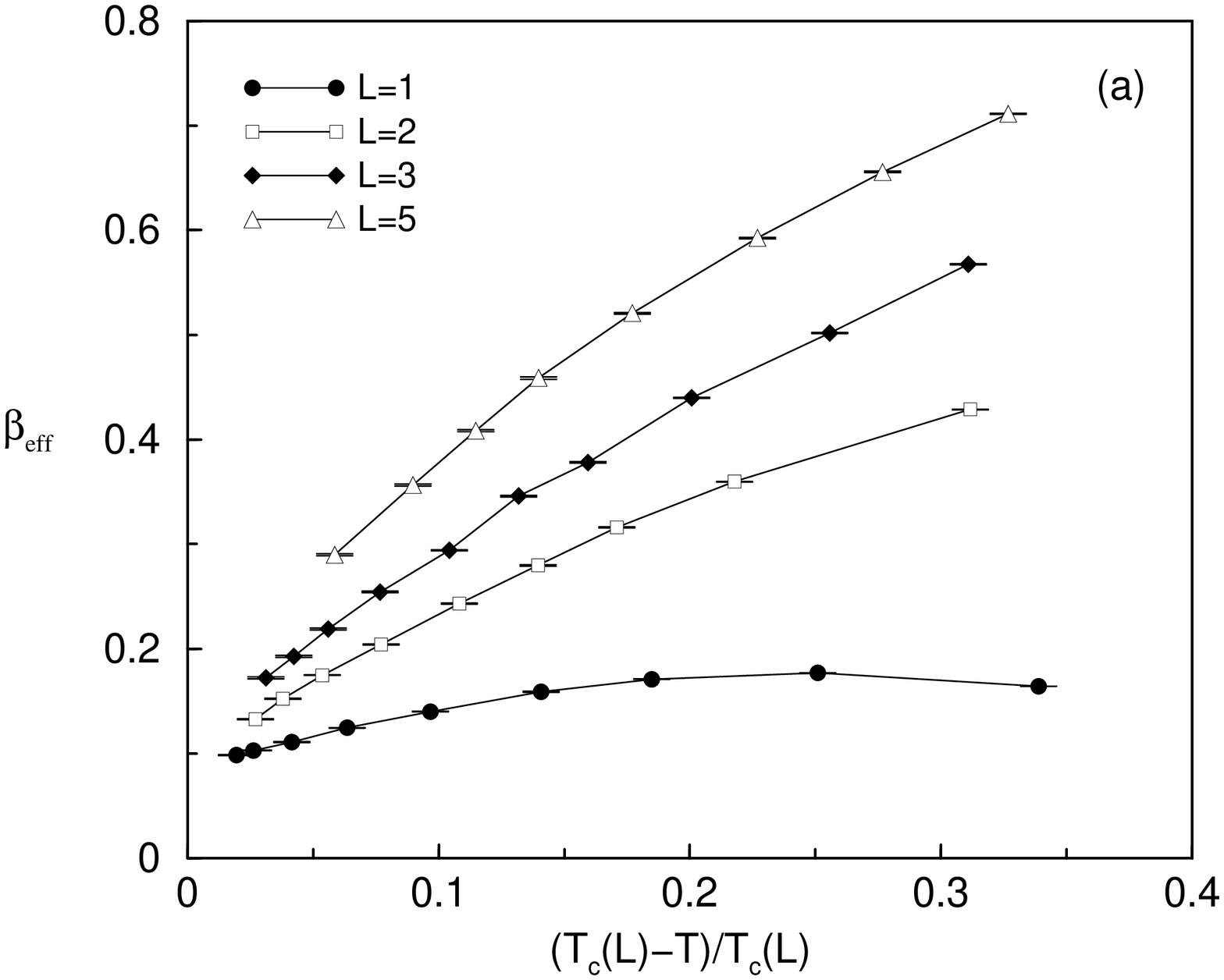,width=3.35in,angle=270}}
\end{figure}
\vspace*{0.2cm}
\begin{figure}
\centerline{\psfig{figure=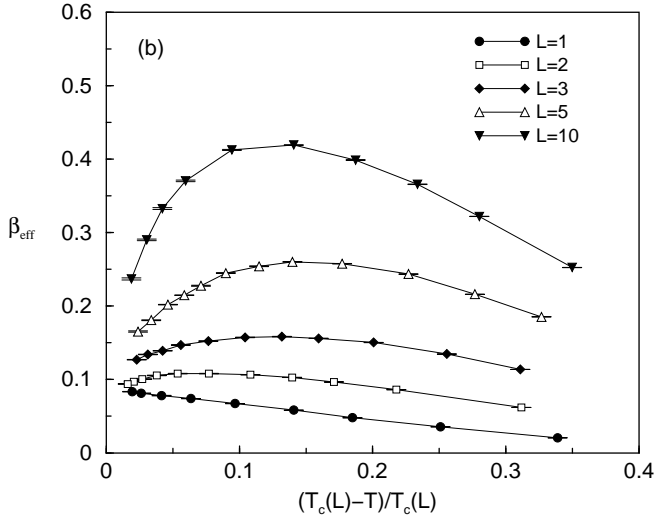,width=3.35in,angle=270}}
\caption{
Effective exponent $\beta_{eff} ((M+1)/2,k;L)$ 
of the magnetization (a) in the additional line of spins ($k= 0$), $m_l$,  
and (b) in the line beneath ($k= 1$), $m_{lb}$, for the Ising model with
equal couplings, as obtained from the MC simulations. The
sizes of the layers are up to $161 \times 320$ spins. Only results which are
essentially free of finite--size effects are shown.}
\label{fig7} \end{figure}
 
(ii) At the {\it surface} transition of flat Ising films, the surface
magnetization vanishes, on
approach to $T_s$, like $m \propto t^{\beta_{2d}}$, independent
of $L$. The critical exponent for the magnetization at the
additional line of magnetic atoms on top of the surface, $\beta_l$, with
$m_{l(lb)} \propto t^{\beta_l}$, depends on the local couplings
at that line, as seen from our DMRG results for $L= 1$. Indeed, the
situation is similar to that of the edge magnetization at the
surface transition, the edge corresponding to an extended defect
line \cite{Ple99,Igl93}, where non--universality holds as well.

For $J_s= 2J_b$ and $J_a= J_n= J_s$, variant A, one obtains for a single
layer, from the DMRG method, $\beta_l(L=1) \approx 0.084$, i.e.\ the value
is below that of the perfect two--dimensional Ising model because of the
increase in $m_l$ due to the
additional line of spins, see Fig.\ 2. The value increases weakly
with layer thickness, becoming in the limit of the semi--infinite
system $\beta_l (L= \infty)= 0.091 \pm 0.002$, as
inferred from MC data for films with thickness $L$ up to 40, and
reasonable extrapolations. Because the critical fluctuations in
a film of finite thickness are ultimatively of two--dimensional
nature, one expects a non--universal critical behaviour at
the defect line with $\beta_l$ depending on $L$. Actually, the slight
increase of the critical exponent with $L$ reflects
the impact of the bulk spins, which now tend to lower
the magnetization in the defect line.

For $J_s= 2J_b$ and $J_a= J_n= J_b$, variant B, both
for single layers, $L=1$, and films, the magnetization
profile $m(i,k;L)$ close to $T_s$ is non--monotonic exhibiting a minimum
at the center line $i= (M+1)/2$, see Fig.\ 8. This minimum is due to
the reduction of the couplings $J_n$ at the defect below the value $J_s$
elsewhere in the surface. As one goes deeper into the bulk, the 
magnetization profile smoothens, which can be readily understood.

The critical exponent describing the vanishing
of $m_l$ (or $m_{lb}$) depends rather
weakly on the thickness $L$ of the film. For $L= 1$, we estimate from
the MC data $\beta_l (L=1)= 0.38 \pm 0.01$, i.e. a value above
the Onsager value of the perfect two--dimensional Ising
model resulting from the decrease of the magnetization at the
defect line \cite{Igl93}. The effective exponent decreases on approach to
criticality, $t \rightarrow 0$, when considering
$m_{lb}$, while it increases when considering $m_l$, allowing
to estimate $\beta_l$ accurately.--From data for fairly
thick films, $L$ up to 40, we
estimate $\beta_l$ of the semi--infinite system to be 
$\beta_l(L= \infty)= 0.34 \pm 0.02$. The slight change of
$\beta_l(L)$ with $L$ for
films of finite thickness is, again, believed to be due to the
correlations of the spins at the defect
line with the bulk spins, which affect $\beta_l$ in such a way that
it is non--universal
when the critical fluctuations are of two--dimensional character.

\begin{figure}
\centerline{\psfig{figure=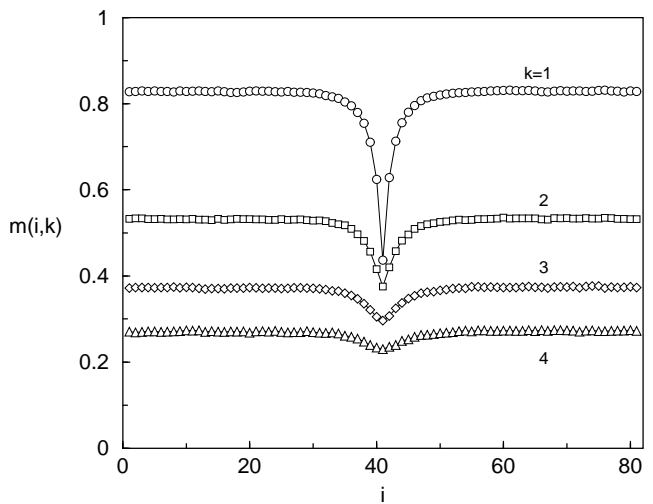,width=3.35in,angle=270}}
\caption{
Magnetization profile $m(i,k;L)$ at different 
depths $k$ in films with $L=40$ layers and one 
additional line of spins on the surface.
The simulation was done for $81 \times 80$ spins in
each layer, couplings $J_s=2 J_b$, $J_a=J_n=J_b$ and
temperature $T=0.96T_c$. The magnetization in
the additional line itself is $m_l \approx 0.218$.}
\label{fig8} \end{figure}
         
\subsection{Step}

Finally, we briefly report our findings for the
critical properties of the step magnetization.
A straight step is introduced (actually two steps, to allow for
periodic boundary conditions) by adding half a layer of magnetic
adatoms to the surface of the magnetic
film \cite{Ple98}, see Fig.\ 1(d). We 
discrimate two couplings, $J_s$ if both neighbouring spins are surface
spins, and $J_b$ otherwise. We consider the line magnetization of
the spins at the step edge, $m_{se}$, vanishing
on approach to the transition as $m_{se} \propto t^{\beta_{se}}$.

For $J_s= J_b$, i.e. at the {\it ordinary} transition, one
obtains $\beta_{se} \approx 0.80$ in semi--infinite Ising
models, $L \rightarrow \infty$, i.e. the same
value as for the critical exponent of the
surface magnetization, as had been shown in
a previous Monte Carlo study on thick
Ising films with a step \cite{Ple98}, in
agreement with analytical considerations \cite{Die90}. However, in thin
films, the critical behaviour is quite different. In the
simulations, for a 
single layer $L=1$ plus half a
layer, we find a critical exponent close to 1/2 (its
concrete value depends rather sensitively on a very accurate 
determination of $T_c$), i.e. a value close to that
of the surface critical exponent $\beta_1$ of the two-dimensional
Ising model (note also its robustness
against randomness in the couplings \cite{Sel97}). This observation
can be understood in
the following way. One is dealing with a composite system
displaying, as the layer size
goes to infinity, two distinct phase transitions, one at the 
critical temperature of the Ising plane, $k_BT_c(L=1)/J_s = 2.269...$,
and one at the critical temperature of the double
layer, $k_BT_c(L=2)/J_s \approx 3.2$. Related composite Ising
models have been investigated before \cite{Bar90,Ber91,Igl90}, showing
that on approach to the upper critical temperature, where
half of the system is disordered, the critical behaviour of
the magnetization at the interface (i.e., here, at the
step) is governed by the surface critical exponent. The same scenario
is expected to hold for finite films with
$T_c(L+1) > T_c(L)$. However, the temperature region where
this behaviour can be observed, will become smaller and smaller as 
$L$ increases. 

At the {\it surface} transition, the same considerations
are believed to be valid. Indeed, in the case $J_s= 2 J_b$, we found 
$\beta_{se}$ to be quite close to 1/2
for a single layer, $L=1$, plus half a layer. For trivial
reasons, $\beta_{se}(L=1)=1/2$ holds for $J_s \gg J_b$, when 
the bottom layer and the extra half layer decouple with the step edge
being the surface of a two--dimensional Ising model. In
the thermodynamic limit, where $T_c(L+1)=T_c(L)$, so that the above
decoupling considerations do not apply, we estimated from MC data for
films with up to 40 layers, a value
of $\beta_{se}= 0.33 \pm 0.02$. Presumably, in
that limit, $\beta_{se}$ is non--universal at the surface
transition, depending on the ratio $J_s/J_b$.

\section{Summary}
Using density-matrix renormalization group and Monte Carlo techniques, we
studied critical properties of magnetic Ising films with various
surface defects. 

In particular, the effect of the local couplings at one or two
additional lines of magnetic adatoms on the surface as well as
at straight steps of monoatomic height has been investigated, especially
in the limiting cases of films consisting of merely one layer and
rather thick films.

In the case of a single layer, $L=1$, with additional
lines of magnetic adatoms, the critical exponent of the
magnetization at the surface defect is non--universal. The dependence
of its value on the 
local couplings, as compared to that of the perfect two--dimensional
situation, follows the trends observed for the exactly soluble
two--dimensional Ising model with ladder and chain like bond--defects.
The value may be lower or larger than in the
perfect situation, 1/8, corresponding to
an increase or decrease in the magnetization at the defect line. Adding
half a layer of spins, one recovers, at the step, the surface critical
exponent, 1/2, of the two--dimensional Ising model.

In the limit $L \rightarrow \infty$, varying the strength
of the surface couplings may lead either
to a surface or an ordinary phase transition. The change
of the critical exponent of the magnetization at the defect has been
found to depend only fairly weakly, for both types of
transition, on the film thickness $L$ in the case of one additional
line of spins. At steps, the critical exponent is argued
to be 1/2, for films of finite thickness and
both kinds of transition, in agreement with the simulations.

In the paper, we have not only presented the results for the 
exponents, but also shown various magnetization curves directly,
so as to give an impression of the size of the effects.
This is meant to encourage further experimental work on such
surface structures and their magnetic properties.

\acknowledgments
We would like to thank K. Baberschke and J. Kirschner
for discussions of the experimental situation. 
M.C.C. thanks the Deutscher Akademischer
Austauschdienst (DAAD) for financial support.

\end{document}